\documentclass[letterpaper]{article}
\usepackage{iccc}

\usepackage{subcaption}
\usepackage{url}
\def\UrlBreaks{\do\/\do-\do_\do.\do=\do@}
\usepackage{times}
\usepackage{helvet}
\usepackage{courier}
\usepackage{booktabs}
\usepackage{array}
\title{I Was Scrolling and Then I Saw a Pregnant Strawberry }
\author{Piera Riccio\\
University of Amsterdam\\
p.riccio@uva.nl
}
\usepackage{booktabs}
\usepackage{colortbl}
\usepackage{xcolor}
\usepackage{rotating}
\usepackage{array}
\definecolor{bodycolor}{RGB}{255, 182, 193}      % pink - Body & Sexuality
\definecolor{violencecolor}{RGB}{255, 160, 122}  % salmon - Violence
\definecolor{betrayalcolor}{RGB}{255, 218, 112}  % yellow - Betrayal
\definecolor{manipcolor}{RGB}{176, 224, 230}     % light blue - Manipulation
\definecolor{moneycolor}{RGB}{152, 251, 152}     % light green - Money & Greed
\definecolor{stigmacolor}{RGB}{216, 191, 216}    % lavender - Stigmatization
\definecolor{cellgray}{RGB}{240,240,240}         % empty cell

\newcommand{\Y}[1]{\cellcolor{#1}}
\newcommand{\N}{\cellcolor{cellgray}}

\begin{document} 
\maketitle
\begin{abstract}
\begin{quote}
AI minidramas (also known as \textit{fruit dramas}) are short, algorithmically distributed generative AI video series featuring anthropomorphized characters that have recently emerged as a widespread phenomenon on social media platforms. This paper argues that despite their seemingly innocuous aesthetic, these videos reproduce deeply gendered narrative structures in which female characters are systematically associated with moral transgression, sexual betrayal, and reproductive capacity, and that several plots also encode the logic of racialization, \textit{i.e., }the process by which visible bodily difference is morally loaded. Drawing on feminist film theory, critical race theory, and platform studies, it further argues that the generative AI aesthetic of these videos, characterized by softness, roundness, and visual cuteness, functions as a mechanism of aesthetic laundering, neutralizing the ideological weight of these narratives and enabling their circulation despite content moderation systems. This paper approaches these questions through personal observation and close reading, reflecting on the specific affordances of generative AI that make this phenomenon both possible and culturally consequential for the field of computational creativity.

\end{quote}
\end{abstract}

\section{Strawberrina}

\begin{figure}[t]
    \centering
    % Verander [b] naar [t]
    \begin{subfigure}[t]{0.41\linewidth}
        \centering
        \vspace{0pt} \includegraphics[width=\linewidth]{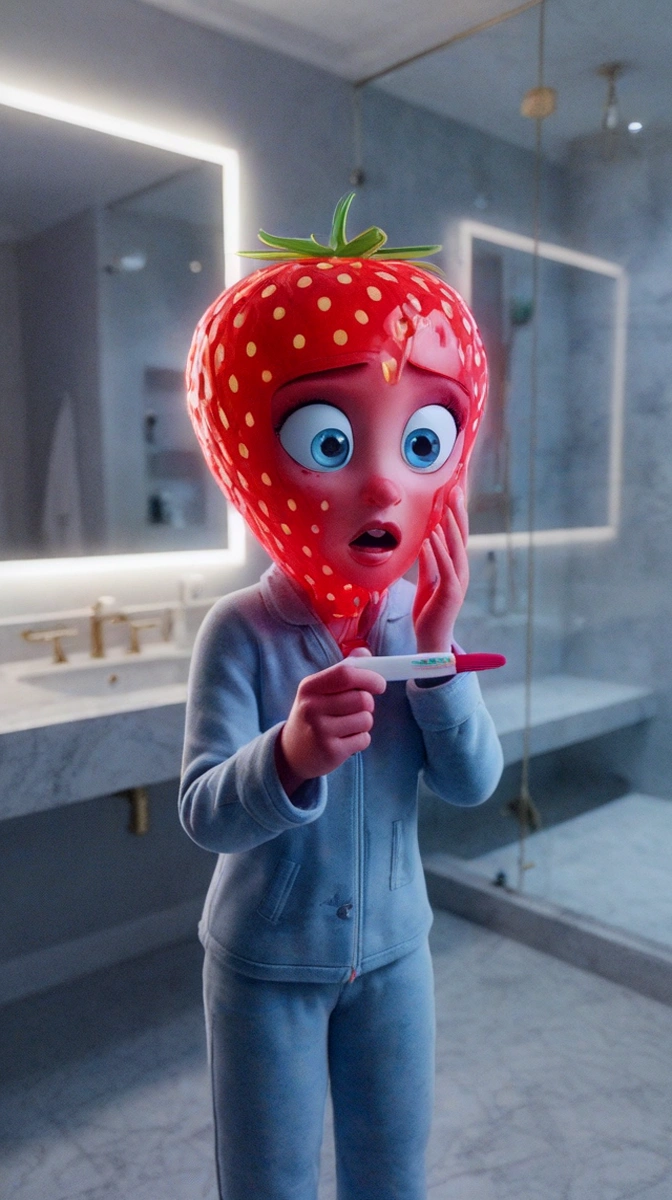}
    \end{subfigure}
    \hfill
    % Verander [b] naar [t]
    \begin{subfigure}[t]{0.54\linewidth}
        \centering
        \vspace{0pt} 
        \includegraphics[width=\linewidth]{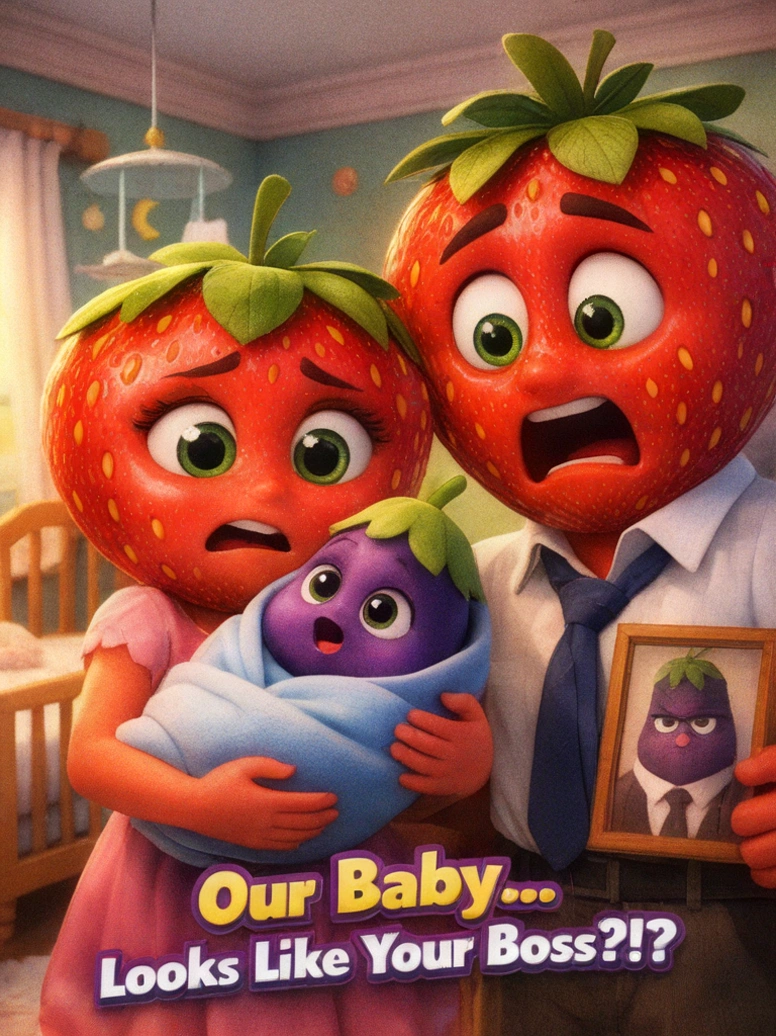}
    \end{subfigure}
    \caption{Screenshots from AI minidramas on TikTok, April 2026. Left: a female fruit character with a pregnancy test. Right: “Our Baby... Looks like your boss?", a couple of Strawberries and their eggplant baby. Original accounts unverified. Both images are exemplary of recurring patterns in the annotated corpus.}
    \label{fig:examples}
\end{figure}

I was scrolling reel after reel on Instagram. I am not proud of this, but I do it when I am bored or when I need a quick distraction. Nothing was quite getting my attention, but I kept scrolling for a while. \textit{I was scrolling and then I saw a pregnant strawberry.} Well, actually, she was some sort of anthropomorphized being. She had arms and legs and a face, she could speak and she was wearing clothes, but her head was shaped like a strawberry and her skin was red (see Figure \ref{fig:examples}). Her name was Strawberrina. She had a cute husband, Strawberrino, and a cute house. They seemed quite happy until Strawberrina gave birth to her child, but the child was an eggplant rather than a strawberry. Strawberrino found out that his wife had been cheating on him with her boss, who was an eggplant, and he felt heart-broken. A mini-drama of around 8 or 9 episodes unfolded, with new episodes being published by the same account daily. Many things happened: a new love for Strawberrino, an orange, from which a new orange-strawberry baby was quickly born (with the doctor being surprised to find out that finally a wife had not cheated on her husband), the baby eggplant got abandoned, Strawberrina tried to rob a bank, she went to jail, and then met a mysterious lemon lawyer that claimed he could bring her out of the trouble.  After this episode, I couldn’t find any follow up on this mini series and the account that was publishing them seemed to disappear from my radar (Figure \ref{fig:examples} shows related examples from other accounts). Because I got interested in this first \textit{fruit drama} that I saw, the recommendation algorithm of Instagram started proposing more and more of such videos to me. Given my academic background in Generative AI, I initially was quite ashamed of having fallen for AI slop and I was feeling \textit{brain rotted} by these nonsensical \textit{fruit dramas}, but soon enough, I felt that this phenomenon deserved my attention actively.

\section{From Brain Rot to Research}

I started noticing some recurring patterns among the videos published by different accounts. The background melancholic music is almost always the same\footnote{for instance: https://www.youtube.com/watch?v=3w7hhqBNoDY, Last Access: 27.04.2026} and the topic across different videos is also quite standardized: love dramas, betrayal, cheating \cite{dazed2026fruit}. Most of the videos included fruits and vegetables, but sometimes the anthropomorphized beings are pets, food, objects, or  anything else we can imagine. The term \textit{fruit drama} \cite{dazed2026fruit} is quite popular, but we generally refer to them as AI minidramas. Such videos represent one of the most widespread and organically emergent instances of generative AI being adopted for creative production by non-expert users at scale \cite{bbc2026fruit}. The reach of this phenomenon extends beyond individual creators: during my observation period, I encountered a \textit{fruit drama} published on the official TikTok account of Lega, the right-wing Italian political party led by Matteo Salvini\footnote{https://vm.tiktok.com/ZNRbpJbvq/, Last Access: 26.04.2026}, demonstrating that the format has been adopted also by institutional political actors as a vehicle for communication and outreach. Understanding what people do with generative AI tools when left to their own devices, \textit{e.g., }what narratives they tell, what aesthetics they converge on, and what cultural values get embedded is a fundamental question for computational creativity as a field \cite{broad2024dead,lopes2024computational}. 

My research focuses on plot structure and character roles, examining how gender and other social constructs are represented in these videos. To analyze this systematically, I manually annotated all AI minidrama videos that appeared on my Instagram feed over more than one month (mid-March to late April 2026), after saving them with the Instagram's native save function. I chose not to download videos out of respect for GDPR and copyright regulations. I annotated whether a main female character was depicted negatively, and why; whether a main male character was depicted negatively, and why; and whether pregnancy was represented, particularly as the result of an extramarital relationship. I annotated 137 videos, in seven languages (English, Spanish, Italian, French, Portuguese, German, and Arabic) or mute, shared by 61 different Instagram accounts. The phenomenon proved unstable during the observation period: several videos and accounts disappeared before annotation was completed, highlighting the volatility of this fast-moving ecosystem. The analyzed corpus is also shaped by Instagram’s recommendation algorithm. As such, these videos are not a random or representative sample of all AI minidramas, but an algorithmically curated selection deemed relevant to a user with my profile, reflecting how most viewers encounter this content.

\section{Gender Roles, Served Fresh}

As anthropomorphized beings (see Figure \ref{fig:examples}), the characters in these AI minidramas have visible gendered features that easily allow distinguishing between binary gender representations (male and female). In most of the videos I have annotated, the male characters represent positive values. Indeed, the most common male character I have observed is a loyal husband, hard working and devoted to his family. The wife is instead often depicted negatively. She can be money-driven, leveraging romantic attachments for financial or social advantage. She is often cheating on her husband, sometimes even with multiple other males. She is also unfaithful to her own friends, jealous and superficial. Only rarely, I have encountered stories in which the husband cheats on the wife, but interestingly enough, in those cases the affair is often the best friend or the sister of the wife. 

\begin{table*}[t]
\centering
\caption{Heatmap of negative female representations in AI minidramas across macro-areas. Numbers refer to categories listed below. Colored cells indicate membership in a macro-area; categories appearing in multiple macro-areas reflect the intersectional nature of these narratives.}
\label{tab:neg}
\small
\setlength{\tabcolsep}{3pt}
\vspace{3mm}
\begin{tabular}{lcccccccccccccccccccccc}
\toprule
& {\textbf{1}}
& {\textbf{2}}
& {\textbf{3}}
& {\textbf{4}}
& {\textbf{5}}
& {\textbf{6}}
& {\textbf{7}}
& {\textbf{8}}
& {\textbf{9}}
& {\textbf{10}}
& {\textbf{11}}
& {\textbf{12}}
& {\textbf{13}}
& {\textbf{14}}
& {\textbf{15}}
& {\textbf{16}}
& {\textbf{17}}
& {\textbf{18}}
& {\textbf{19}}
& {\textbf{20}}
& {\textbf{21}}
& {\textbf{22}} \\
\midrule
\textbf{Body \& Sex}      
& \N & \N & \Y{bodycolor} & \Y{bodycolor} & \N & \N & \N & \Y{bodycolor} & \N & \N & \Y{bodycolor} & \N & \N & \Y{bodycolor} & \N & \N & \N & \Y{bodycolor} & \N & \Y{bodycolor} & \Y{bodycolor} & \Y{bodycolor} \\
\textbf{Violence}          
& \Y{violencecolor} & \N & \N & \N & \Y{violencecolor} & \N & \N & \N & \N & \N & \N & \N & \N & \N & \N & \Y{violencecolor} & \N & \N & \Y{violencecolor} & \N & \Y{violencecolor} & \N \\
\textbf{Betrayal}          
& \N & \Y{betrayalcolor} & \N & \Y{betrayalcolor} & \N & \N & \N & \N & \N & \Y{betrayalcolor} & \N & \N & \N & \N & \N & \N & \N & \N & \Y{betrayalcolor} & \N & \N & \N \\
\textbf{Manipulation}      
& \N & \N & \N & \N & \N & \N & \N & \Y{manipcolor} & \Y{manipcolor} & \N & \Y{manipcolor} & \N & \Y{manipcolor} & \N & \N & \N & \N & \N & \N & \N & \N & \N \\
\textbf{Money \& Greed}    
& \N & \N & \N & \N & \N & \Y{moneycolor} & \Y{moneycolor} & \N & \N & \N & \N & \Y{moneycolor} & \Y{moneycolor} & \N & \N & \N & \Y{moneycolor} & \N & \N & \N & \N & \N \\
\textbf{Stigmatization}    
& \N & \N & \Y{stigmacolor} & \N & \N & \N & \N & \N & \N & \N & \N & \N & \N & \Y{stigmacolor} & \Y{stigmacolor} & \N & \N & \N & \N & \N & \N & \N \\
\bottomrule
\end{tabular}

\vspace{0.5em}
\noindent\small\textbf{Category legend:} 
1.~Being a bully; 
2.~Being disloyal to friends/family; 
3.~Body stigmatization; 
4.~Cheating; 
5.~Child abuse; 
6.~Corruption; 
7.~Criminality; 
8.~Deceptive intimacy; 
9.~Emotional detachment; 
10.~Envy; 
11.~False victimhood; 
12.~Financial parasiting; 
13.~Gaslighting for personal benefit; 
14.~Honor-based shame; 
15.~Intellectual stigmatization; 
16.~Maternal abandonment; 
17.~Obsession for money; 
18.~Obsession for sex; 
19.~Revenge; 
20.~Sex for career advancement; 
21.~Sexual violence; 
22.~Spreading STDs.
\end{table*}

While not every female character is depicted negatively, the “evil” in most of the minidramas I have watched is depicted through a \textit{main} female character (in 73.7\% of cases, vs. only 33.6\% of main male characters depicted negatively). For example, in the first story that I described, Strawberrina cheated on her husband with her eggplant boss. Arguably, the boss of Strawberrina is making an immoral action by having a sexual relationship with one of his employees. However, the narration steers away from the boss and focuses on Strawberrina not being a loyal wife. Having a narration that unfolds around the (negative) actions of a female character means that these characters have a narrative agency and they are not passive recipients of the actions of the males around them. In contrast, most of the male characters are not represented as “heroes”, but more as “victims”, placing them in a more passive and flat representation style compared to the female counterparts. Importantly, male characters may appear more naive and passive, but they are naive and passive in a way that leaves their fundamental goodness intact. While her activity is her corruption, his passivity is his innocence. 

Analyzing my annotations, I have summarized the negative representations of female characters into 22 categories (reported in Table~\ref{tab:neg}) and further grouped into 6 macro-areas (Body \& Sex, Violence, Betrayal, Manipulation, Money \& Greed, Stigmatization). Among these 22 categories, \textit{Body Stigmatization}, and \textit{Intellectual Stigmatization} are notable because they do not depict the female character as actively doing something wrong. Rather, she is stigmatized for conditions or choices that are neutral or entirely natural: her body shape, her eating habits, bodily functions such as flatulence or menstruation, or she is depicted as intellectually inferior. In \textit{Honor-Based Shaming}, the female character is shamed not directly for her own sake, but because her choices, such as having an account on OnlyFruits (the AI minidrama equivalent of OnlyFans), bring social embarrassment to the male members of her family. Her body and her sexuality are treated as property of the males around her, and her autonomous choices become their problem. The ideological work of these categories is precisely to present neutral or natural features as sources of moral failing, which is itself a form of misogyny. In addition, the category of \textit{Deceptive intimacy }includes narratives in which a female accepts payment for a sustained years-long relationship during which she simulates love, builds a family, and performs domesticity, only to disappear with the money once the agreed term expires. The category of \textit{False Victimhood} features scenarios in which a female character sexually assaults a male and then, exploiting her social position as a female, convinces those around her that she was the victim. This narrative is particularly harmful because it directly reproduces one of the most persistent and damaging myths used to discredit women affected by sexual violence \cite{burt1980cultural}. The category of \textit{Financial Parasiting} and \textit{Sex for Career Advancement } similarly frame female economic agency as inherently predatory: the female who seeks financial security or professional advancement is depicted not as navigating structural inequalities but as a manipulative agent exploiting male vulnerability. 

Taken together, the analyzed categories suggest that what is most threatening to the narrative logic of AI minidramas is not female evil per se, but female autonomy (economic, sexual, and social) which is systematically reframed as corruption. The figure of the “evil woman" as a site of displaced agency is something feminist film scholars have been writing about for decades. The classic example is the \textit{femme fatale} \cite{doane2013femmes}, a character who is morally coded as dangerous and destructive, almost always in relation to sexuality, but who also has a kind of narrative power and centrality that the “good" female characters completely lack \cite{mulvey2013visual}. However, while the evil female character is central in these minidramas, she is central because of her relationship to reproduction, sexuality, and betrayal. Her agency is granted on precisely the terms that have historically been used to control and pathologize women. This tension can be related to the concept of “postfeminist masquerade" \cite{mcrobbie2008aftermath}, \textit{e.g.}, the way certain representations seem to give women power and centrality while simultaneously reinforcing the terms of their subordination. 
The female character drives the plot, yes, but the plot she drives is one entirely organized around her body, her reproductive capacity, and her sexual transgression. Pregnancy, in particular, is quite represented in these minidramas (appearing in 31.4\% of my annotated videos) and is almost depicted as a physical condition that women, as such, cannot avoid and that defines the very essence of being a female in these plots. Pregnancy itself becomes a symbol through which the storytelling revolves around the body of the female characters rather than their (negative) actions or values.

\section{The Eggplant in the Room}

When Strawberrina gives birth to her child, both the doctor and the husband realize that she had cheated because her baby looks “different". In human life, this could only happen when the baby's physical appearance, particularly skin color or other visible traits, makes it immediately obvious that the biological father is someone else. This is a quite common dynamic across my annotated videos (65.1\% of videos showing pregnancy), and I find it one of the most uncomfortable things to sit with analytically. The choice of making characters different species of fruit, vegetables or objects is doing something quite specific and not at all innocent: the “wrongness" of the baby is legible at a glance because it is different from the family it was born into. Such narratives not only reproduce misogynistic gender dynamics, but they also encode  what scholars of critical race theory would call the logic of racialization \cite{omi2014racial}: the process by which visible bodily difference becomes meaningful, morally loaded, and socially consequential. As previously mentioned, these AI minidramas are circulating globally, across audiences with very different cultural frameworks and histories, where the meaning of this narrative may shift and mutate for different audiences \cite{doh2025position}. This does not make the structural observation less valid, but it does mean that a single universal reading should not be assumed. Despite this, the narrative structure of these videos relies on the logic of racialized bodily difference, with the idea that you can look at a body and immediately know where it does and does not “belong” \cite{fanon2016black,chi_workshop}. The anxiety about paternity that runs through these minidramas is not a neutral dramatic device: it is historically entangled with fears about racial mixing, bodily purity, and the integrity of the family as a unit of racial reproduction. Indeed, reproductive capacity has historically been weaponized against racialized women specifically, with their bodies treated as sites of anxiety, surveillance, and control \cite{roberts2014killing}. While the eggplant baby of Strawberrina is visually funny, the story it echoes is not. 

In addition, across several videos in my corpus, I observed characters actively concealing their species or color identity in order to access better social status, a structural parallel to the phenomenon of racial passing \cite{fanon2016black} also reproducing logics of colorism \cite{hunter2007persistent}: \textit{e.g.,} green apples over red apples, or precious stones ranked by their market value as proxies for social worth.  In one particularly striking example, a brown strawberry had spent his entire life painting himself red to pass as a red strawberry, only to be exposed when his twins were born (one red, one brown) and he and his wife proceeded to systematically discriminate against the brown child. These are not incidental narrative details: the fruit world, it turns out, has a “racial" order. 

\section{Aestheticized Violence}

But if such AI minidramas encode misogynist and racist values, one could wonder how it is possible that they are becoming so viral, reaching such a variety of audiences (including myself). It is exactly the circulability and platformization of this content that make AI minidramas particularly insidious. Many of these videos depict images that are not light, including forced pregnancies, sexual assault, child abuse, physical violence, but they present a cute visual aesthetics that I frame here as a form of \textit{aesthetic laundering} \cite{downey2024aesthetic}, a process by which disturbing content presented as innocuous through its visual register. The cuteness of the videos and the AI-generated anthropomorphized bodies create a visual register that neutralizes the weight of what is actually being depicted. In other visual genres, such topics would demand a certain treatment: slowness, gravity, consequence, space for the viewer to process. But wrapped in pastel colors and round unrealistic bodies, moving at the pace of a short-form reel, they just... pass. The aesthetic does ideological work by making the unbearable feel inconsequential. Cuteness, in these videos, can be seen as a mode of power and disavowal and it is, indeed, an aesthetic category that is not innocent \cite{ngai2010our}: it manages and contains affect, it makes things seem harmless and manipulable precisely because they are rendered small and soft and round. Crucially, cuteness is also an economic engine. Characters that seem to have souls and affective depth are precisely designed to generate attachment that can be monetized \cite{allison2006millennial}. The AI aesthetic carries this cuteness, operating within neoliberal logics of attention and profit, with their seemingly innocuous aesthetics not appearing serious enough to be suspicious \cite{banet2018empowered}. Yet AI minidramas are distinct from earlier cute commodification, first, because of the simplicity to generate several instances of such content \cite{slop_infrastructures}, and second because the cute aesthetic is not solely a deliberate creative choice and it is partly an artifact of the generative models themselves, which encode and amplify existing aesthetic and social norms through their training data \cite{jaaskelainen2025intersectional}. 

In addition, these videos circulate on online platforms where normally content moderation practices remove or shadow ban content that is considered not-safe-for-work (NSFW), effectively influencing the contemporary creative ecosystem \cite{riccio2022algorithmic}. However, the definition of “safety" in these systems is quite controversial and ambiguous \cite{gillespie2018custodians}, often prioritizing content that generates revenues and engagement \cite{riccio2024exposed}. Indeed, the AI minidramas effectively bypass content moderation filters, depicting images that would normally be filtered out of these platforms. And what passes moderation gets recommended. Platform recommendation algorithms reward engagement, and the melodramatic structure of these minidramas (the cliffhangers, the betrayals, the shocking reveals) is also precisely engineered for it. The cuteness gets the content through the filter; the drama keeps the viewer watching; and the algorithm reads both signals as reasons to push the video further, rather than restricting it. Content moderation and content recommendation are, in this sense, two faces of the same market-driven logic of these platforms \cite{riccio2022algorithmic}, and the AI minidrama aesthetic has learned to exploit both simultaneously. 

\section{Delusions on Creativity}

While the creators of visual generative AI models often propose revolutionary creative possibilities of these technological tools \cite{epstein2023art}, AI minidramas, being a mass-scale, real-world instance of generative AI being used creatively by non-expert users, reveal that computational creativity in practice is not neutral. The creative affordances of these tools (what they make easy, what they make possible, what they make look “good") are shaping cultural output in ways that have ideological consequences.
These are not merely aesthetic or media culture questions. If generative AI tools systematically make certain visual styles easier, certain narrative structures more template-ready, and certain ideological contents more platform-compatible, then the creativity they enable is already shaped before the creator makes a single choice. The non-expert user picking up these tools is not a blank creative agent: they are working within a sociotechnical assemblage \cite{stivale2014gilles,delanda2019new} that has its own affordances, its own defaults, and its own politics. This assemblage includes the generative models and their training data, the platforms and their recommendation and moderation logics, the creators and their prompting strategies, and the audiences and their engagement behaviors, and it is where computational creativity actually happens in the wild. And it is at the level of this assemblage, not of any single tool or actor, that ideology is produced and distributed.
What AI minidramas reveal, perhaps more clearly than any controlled creative experiment could, is that computational creativity in the wild is never just about what is made. It is about what the tools make easy, what the platforms make visible, and what the aesthetics make bearable. 

Strawberrina was generated by someone, recommended to me, and eventually became the unlikely muse of this short paper. That chain, going from model to platform to creator to algorithm to audience to researcher, is where the ideology lives. The field of computational creativity has invested enormous effort in understanding what computational systems can make. The present paper suggests it is equally urgent to study what they make routine, what narratives they normalize, what aesthetics they reward, and what societal representations they quietly distribute at scale.

\section{Acknowledgments}
I would like to thank Miriam Doh, Francesco Galati, Noa Garcia, Benedikt Höltgen, Ludovica Schaerf, Carla Teodoro Cubeñas, and Nanne van Noord for supporting this idea and for providing insightful feedback on this paper.

\bibliographystyle{iccc}
\bibliography{iccc}

\end{document}